\documentclass[journal]{IEEEtran}

\usepackage{graphicx}
\DeclareGraphicsExtensions{.pdf,.jpeg,.jpg,.png}

\ifCLASSOPTIONcompsoc
    \usepackage[caption=false, font=normalsize, labelfont=sf, textfont=sf]{subfig}
\else
\usepackage[caption=false, font=footnotesize]{subfig}
\fi 

\usepackage{amsmath}
\usepackage{amssymb}
\usepackage{physics}
\usepackage{xcolor}
\usepackage[super]{nth}
\usepackage{algorithm}
\usepackage{algpseudocode}

\usepackage{cases}
\usepackage{cite}
\usepackage{threeparttable}

\newcommand{\etal}{\textit{et al}. }

\begin{document}
\title{Dynamic Power Systems Line Outage Detection Using Particle Filter and Partially Observed States}

\author{Xiaozhou~Yang,~\IEEEmembership{Student Member,~IEEE}%
,~Nan~Chen,~\IEEEmembership{Member,~IEEE}
,~and~Chao~Zhai,~\IEEEmembership{Member,~IEEE}

\thanks{X. Yang and Nan Chen are with the Department of Industrial Systems Engineering and Management, National University of Singapore, Singapore 117576, Singapore. X. Yang is also with ETH Zurich, Future Resilient Systems, Singapore-ETH Centre, 1 CREATE Way, CREATE Tower, 138602 Singapore.}%
\thanks{Chao Zhai is with the School of Automation, China University of Geosciences, Wuhan 430074, China, and also with the Hubei Key Laboratory of Advanced Control and Intelligent Automation for Complex Systems, Wuhan 430074,
China. To whom correspondence may be addressed: isecn@nus.edu.sg or zhaichao@amss.ac.cn.}%
}

\maketitle

\begin{abstract}
Real-time transmission line outage detection is difficult because of partial phasor measurement unit (PMU) deployment and varying outage signal strength. 
Existing detection approaches focus on monitoring PMU-measured nodal algebraic states, i.e., voltage phase angle and magnitude. The success of such approaches, however, is largely predicated on strong outage signals and the presence of PMUs in the outage location's vicinity. 
To overcome these limitations, a unified framework is proposed in this work by utilizing both nodal voltage information and generator dynamic states, e.g., rotor angular position. The proposed scheme is shown to be faster and more robust to unknown outage locations through the incorporation of generator dynamics. Using the IEEE 39-bus system simulation data, the proposed scheme's properties and performances compared to existing approaches are presented. 
The new approach could help improve operators' real-time situational awareness by detecting outages faster and providing a breakdown of outage signals for diagnostic purposes, making future power systems more resilient.

\end{abstract}


\section{Introduction}
\IEEEPARstart{P}{ower} systems experience numerous types of disruptions; transmission line outage, in particular, is extensively studied due to its frequent occurrence and potentially severe consequence. Outages can happen due to system faults, power grid component degradation, adverse weather conditions, or vandalism. Detecting events such as line outages is a crucial aspect of real-time situation awareness in the power system control room. Such detection capability enables independent system operators to respond to abnormal events promptly, preventing possibly cascading failures.

The popularization of Phasor Measurement Units (PMUs) makes real-time power system analytics possible. PMUs are devices installed at substations (buses) capable of recording high-fidelity GPS time-synchronized phasors, i.e., physical quantities with both magnitude and phase. Much work has been done recently by utilizing the high-reporting rate of PMUs and, in some approaches, augmenting it with first-principle models to detect outages. There are two challenges to an effective detection scheme. Firstly, outages at lines with relatively small power flow tend to create minimal and short-lived disruptions, e.g., in voltage phase angle. For these outages, detecting them quickly is difficult due to a large signal-to-noise ratio of the measurements. Secondly, only a limited number of PMUs are deployed in a network due to economic and engineering feasibility constraints, leaving some buses unobservable. Therefore, the allocation of PMUs, in terms of the number and location, impacts the scheme's effectiveness. It remains a challenge to design a detection scheme robust to the location of the PMUs and outages.

\section{Related Works}
With the above challenges, it may be advantageous to exploit system transient dynamics following an outage. Transient dynamics are the evolution of the system algebraic state variables, e.g., bus voltage phase angle and magnitude, and dynamic state variables, e.g., generator rotor speed and angle, between two quasi-steady states. It is the result of the synchronous machines reacting to the power imbalance created by the outage. These dynamics provide direct and accurate characterizations of the network's disruption. 

Most state-of-the-art works on outage detection can be grouped by the type of dynamics considered in their formulations. The first group models power systems based on the quasi-steady state assumption where no dynamics are considered \cite{Tate2008, Chen2016,Ardakanian2019a}. However, transient dynamics can often last up to several seconds and are non-negligible. Therefore, this approach may not be adequate at describing the actual system behavior at a fine time scale. The second group relaxes the quasi-steady state assumption and attempts to account for the post-outage transient dynamics. Using time-dependent participation factor matrices, Rovatsos $\etal$ modeled the evolution of voltage phase angles \cite{Rovatsos2017}. Similarly, focusing on voltage angles, a time-varying nonlinear relationship was derived from the alternating current (AC) power flow model and served as the detection basis \cite{yang2020control}. Dynamics are also approximated using low-dimensional subspaces derived from PMU measurements using, for example, principal component analysis (PCA) \cite{Xie2014}, and hidden Markov model (HMM) \cite{Huang2016b}. Correlation matrices obtained from algebraic state variables are also used to uncover underlying changing relationships, e.g. from adjacent bus voltage and current phasors \cite{Jamei2017a} and from the observation matrix obtained during outage-free operation \cite{Hosur2019}. These methods' use of the AC power flow model can describe system dynamics more accurately. The introduction of time-dependent relationships is also a worthwhile attempt at capturing system disturbances. However, they rely solely on algebraic state variables, e.g., bus voltage and current. Generators are the sources of transient dynamics; their internal dynamic states can better characterize the system's transient response to the power imbalance created by the outage.

The third group models the power system as a dynamic system, utilizing both the measurable algebraic states and unobservable generator dynamic states. Using the swing equation, Pan $\etal$ formulated outage diagnosis as a sparse recovery problem solved by an optimization algorithm \cite{Pan2015a}. Similarly, using the swing equation, a visual observer network is constructed to monitor line admittance changes by a parameter identification method \cite{Yang2016b}. Both works focus on the outage diagnosis problem, i.e., line localization and parameter estimation, rather than the detection problem. They also assume all buses are equipped with PMUs that the system is completely observable. Therefore, systematic outage detection using both dynamic and algebraic state information under a partial PMU deployment remains an unsolved problem.

In general, statistical monitoring of dynamic systems, e.g., power systems, is mainly investigated through two approaches. One is the model-free approach, where no knowledge about the underlying process is assumed. Various data-driven subspace identification methods are employed to track system states, e.g., using PCA \cite{choi2004nonlinear}. On the other hand, when an analytical model that can sufficiently characterize the system's dynamic behavior is available, a model-based approach is usually considered. This approach builds on state estimation through various filtering techniques, such as Kalman filters \cite{Zhao2017} and particle filters (PFs) \cite{Cui2015}. Monitoring schemes are then formulated using signals defined by Kalman innovation vectors, i.e., residual monitoring \cite{chen2009detectability}, or measurement likelihoods in the PF's case \cite{Kadirkamanathan2002}.

There is limited work on line outage detection considering generator dynamics in a partially observed network to the best of the authors' knowledge. No work brings together dynamic and algebraic state information for outage detection. In this work, we track system transient dynamics through nonlinear state estimation via a PF. A statistical change detection scheme is constructed by monitoring the PF-predicted output's compatibility with the expected outage-free measurement. When an outage happens, a significant reduction in the compatibility triggers an outage alarm. This work has three main contributions: 1) A novel unified detection framework that incorporates both generator bus dynamics and load bus power changes is proposed. 2) The framework also facilitates post-outage diagnostic work through a breakdown of the monitored system signals. 3) Extensive simulation studies on IEEE test system demonstrate the robustness of the proposed framework against partial PMU deployment and the effectiveness against other state-of-the-art outage detection methods. 

The rest of this paper is organized as follows. A unified outage detection scheme based on nonlinear power system dynamics is formulated in Section \ref{sec:power_model} and Section \ref{sec:detection_scheme}. Section \ref{sec:state_estimation} describes the PF-based online state estimation necessary for tracking generator dynamics. The proposed scheme's effectiveness and advantages are presented in Section \ref{sec:results} using simulation studies. Section \ref{sec:conclusion} is the conclusion.

\section{Power System State-Space Modelling}
\label{sec:power_model}
In this section, we detail a power system model that captures both the generator dynamics and load bus power flow information in a unified framework. Consider a power system with $M$ generator buses where $\mathcal{N}_g = \{1, \dots, M\}$, $N-M$ load buses where $\mathcal{N}_l = \{M+1, \dots, N\}$, and $L$ transmission lines where $\mathcal{L} = \{1, \dots, L\}$. 
The power system is a hybrid dynamical system described by a differential-algebraic model. The second-order generator model, also known as the swing equation \cite{Kundur1994}, is used in this work\footnote{Although the swing equation is used here to model generator rotor dynamics, high-order and more complex models, such as the two-axis model, can be used to develop the detection scheme proposed in this work similarly.}.
For every generator bus $i \in \mathcal{N}_g$, their states are modeled as the differential variables, i.e., $\boldsymbol{X}= [\delta, \omega]^T$ where $\delta$ is the rotor angular position in radians with respect to a synchronously rotating reference, and $\omega$ is the rotor angular velocity in radians/second. The differential equations governing their dynamics are
\begin{subequations}
\label{eqn:de_continuous}
\begin{align}
\dot{\delta}_{i} &=\omega_s\left(\omega_{i}-1\right) \,, \label{ch4:eqn:transition_function_delta}\\
M_i \dot{\omega}_{i} &=\text{P}_{m, i}-\hat{\text{P}}_{g, i}-D_i\left(\omega_{i}-1\right) \,,
\label{eqn:transition_function_omega}
\end{align}
\end{subequations}
where $\dot{\delta}_{i}$ is the derivative of $\delta_i$ with respect to $t$. $\omega_s$ is the synchronous rotor angular velocity such that $\omega_s= 2 \pi f_{0}$ where $f_0$ is the known synchronous frequency. $\text{P}_{m,i}, M_i$, and $D_i$ denote the mechanical power input, the inertia constant, and the damping factor, respectively. They are assumed known and constant for the duration of this study. The inputs for the model are the generated active power, i.e., ${u}= \text{P}_{g}$. Under the classical generator model assumptions, the synchronous machine is represented by a constant internal voltage of magnitude $\text{E}$ and angle $\delta$ behind its direct axis transient reactance $\text{X}_{{d}}^{\prime}$ \cite{Kundur1994}. Therefore, the active power at generator $i$ is
\begin{equation}
\label{eqn:ae_continuous}
\text{P}_{g, i} = \frac{\text{E}_i\text{V}_i}{\text{X}_{d, i}^{'}}\sin (\delta_i - \theta_i)\,,
\end{equation}
where $\theta_i$ is the generator bus nodal voltage phase angle. The transient reactance $\text{X}_{d, i}^{'}$ is assumed known and constant, whereas a method will be presented later to adaptively infer the parameter $\text{E}$ with online data. Also, denote 
$
\hat{\text{P}}_{g, i} = \text{P}_{g, i} + \epsilon_i \,,
$
where $\epsilon$ is assumed to be a zero-mean Gaussian process with a known variance representing the random fluctuations in electricity load on the bus as well as process noise. 

The outputs of the system model are nodal voltage magnitudes and phase angles which PMUs can measure. More importantly, the algebraic output and generator dynamic states have to satisfy an active power balance constraint. The constraint stipulates that the net active power at a bus is the difference between the active power supplied to it by the generator and the load consumed, i.e.,
\begin{equation}
\label{eqn:power_balance}
\text{P}_i = \text{P}_{g, i} - \text{P}_{l, i} \,,
\end{equation}
for $i = 1, \dots, N$, subject to random demand fluctuations $\epsilon_i$ as mentioned above. $\text{P}_{l, i}$ is the load bus $i$, $\text{P}_{i}$ is the nodal net active power and 
\begin{equation}
\label{eqn:ac_pf}
\text{P}_{i} = \text{V}_i \sum_{j=1}^{N} \text{V}_j \text{Y}_{ij} \cos (\theta_i - \theta_j - \alpha_{ij}) \,,
\end{equation} following the alternating current (AC) power flow equation where $\text{Y}_{ij}e^{j\alpha_{ij}}$ are elements of the bus admittance matrix. Note that for load buses $\text{P}_{g, i} = 0$ in (\ref{eqn:power_balance}). The total active power generated and load demand of the network are assumed to be balanced as well. This relationship will be the basis for our unified outage detection scheme described in the next section. 

We define the discrete counterparts of the system model via a first-order difference discretization by Euler's formula, i.e., let $\delta_{k+1} = \delta_{t_{k+1}}$ for $k = 1, 2, \dots$, and $\dot{\delta}_{t_{k+1}} \approx (\delta_{k+1}-\delta_{k})/\Delta t$. For PMU devices with a sampling frequency of 30 Hz, $\Delta t = t_{k+1} - t_{k} = 1/30$ s. Thus, the continuous system of a generator bus $i$ can be approximated by
\begin{equation}
\label{eqn:de_discrete}
\boldsymbol{X}_{i,k+1} = 
\left[
\begin{array}{c}
\delta_{i,k+1} \\
\omega_{i,k+1}
\end{array}
\right] =
\left[
\begin{array}{c}
\delta_{i,k} + \Delta t \omega_s\left(\omega_{i,k}-1\right) \\
\omega_{i,k} + \frac{\Delta t}{M_i}q_{i, k} - \epsilon_{k}\,
\end{array}
\right]  
\end{equation}
where $q_{i, k} = \text{P}_{m, i}-\text{P}_{g, i,k}-D_i\left(\omega_{i,k}-1\right)
$ for notational brevity, and 
\begin{equation}
\label{eqn:p_g_discrete}
\text{P}_{g,i,k} = \frac{\text{E}_i\text{V}_{i, k}}{\text{X}_{d, i}^{'}}\sin (\delta_{i, k} - \theta_{i,k}) \,.
\end{equation}

For the power balance equation (\ref{eqn:power_balance}), we can take a derivative with respect to time $t$ on both sides to obtain \begin{equation}
  \frac{\partial \text{P}_i}{\partial t}  = \frac{\partial \text{P}_{g, i}}{\partial t} -  \frac{\partial \text{P}_{l, i}}{\partial t}  \,.
\end{equation} 
The discretized relationship becomes
\begin{equation}
\label{eqn:ae_discrete}
\Delta\text{P}_{i, k} = \Delta\text{P}_{g, i, k} -  \Delta\text{P}_{l, i, k}\,,
\end{equation}
where $\Delta\text{P}_{i, k} = \text{P}_{i, k} - \text{P}_{i, k-1}$ and similarly for the other two terms. Writing the whole system in vector form, we also define the N-dimensional output variable
\begin{equation}
\label{eqn:ae_discrete}
\boldsymbol{Y}_{k} = \Delta\textbf{P}_{k} =
\Delta\textbf{P}_{g, k}  - \Delta\textbf{P}_{l, k}\,,
\end{equation}
where $\Delta\textbf{P}_{l, k}$ represents the random load fluctuations. $\Delta\textbf{P}_{l, k}$ is assumed to be a zero-mean Gaussian variable with covariance $\sigma^2\mathbf{I}$. Note that the entries corresponding to load buses in $\Delta\textbf{P}_{g, k}$ are all zero. 

Through (\ref{eqn:ae_discrete}), the active power changes in both generator and load buses can be monitored. In comparison, detection schemes developed in previous works focus on monitoring changes in net active power, $\Delta\textbf{P}$, through direct current (DC), e.g., \cite{Chen2016}, or AC, e.g., \cite{yang2020control}, power flow equations. Their formulations can be considered as special cases of the proposed unified framework when no generator information is available, e.g., no PMUs are installed on generator buses. However, as shown in simulation studies, having generator power output information helps to detect certain outages when net active power changes are not significant enough to trigger an alarm.

Equations (\ref{eqn:de_discrete})-(\ref{eqn:ae_discrete}) define a state-space model (SSM) for the power system that could be summarized in a general form:
\begin{subequations}
\label{eqn:general_ssm}
\begin{align}
\boldsymbol{X}_{k+1} &= a(\boldsymbol{X}_{k}, \boldsymbol{u}_{k}, \boldsymbol{\epsilon}_{k}) \, \rightarrow f(\boldsymbol{X}_{k+1}|\boldsymbol{X}_{k}) \\ 
\boldsymbol{Y}_{k} &= b(\boldsymbol{X}_{k}, \boldsymbol{u}_{k}, \boldsymbol{\eta}_{k}) \, \rightarrow g(\boldsymbol{Y}_{k} | \boldsymbol{X}_{k})
\end{align}
\end{subequations}
The dynamics of the unobservable generator states $\boldsymbol{X}$ are governed by the state transition function $a(\cdot)$ as in (\ref{eqn:de_discrete}). The output $\boldsymbol{Y}$, computable from PMU measurements, is governed by the output function $b(\cdot)$ as in (\ref{eqn:ae_discrete}). Since $b(\cdot)$ is a nonlinear function of the system states, the power system is a nonlinear dynamic system. As the transition process is stochastic due to random load fluctuations and measurement errors, the states and outputs can be expressed in a probabilistic way. In particular, we denote the state transition density and output density as $f(\boldsymbol{X}_{k+1}|\boldsymbol{X}_{k}=\boldsymbol{x}_{k})$ and $g(\boldsymbol{Y}_{k} |\boldsymbol{X}_{k}=\boldsymbol{x}_{k})$, respectively, where $f(\cdot)$ and $g(\cdot)$ are probability density functions (PDFs). These two densities play important roles in the particle filter-based state estimaton described in later in the article. An important consequence of the SSM is the conditional independence of the states and output due to the Markovian structure. In particular, given $\boldsymbol{X}_{k}$, $\boldsymbol{X}_{k+1}$ is independent of all other previous states; similarly given $\boldsymbol{X}_{k}$, $\boldsymbol{Y}_{k}$ is independent of all other previous states.

\section{EWMA-Based Outage Detection Scheme}
\label{sec:detection_scheme}
\begin{figure}[!t]
\centering
\includegraphics[width=1\linewidth]{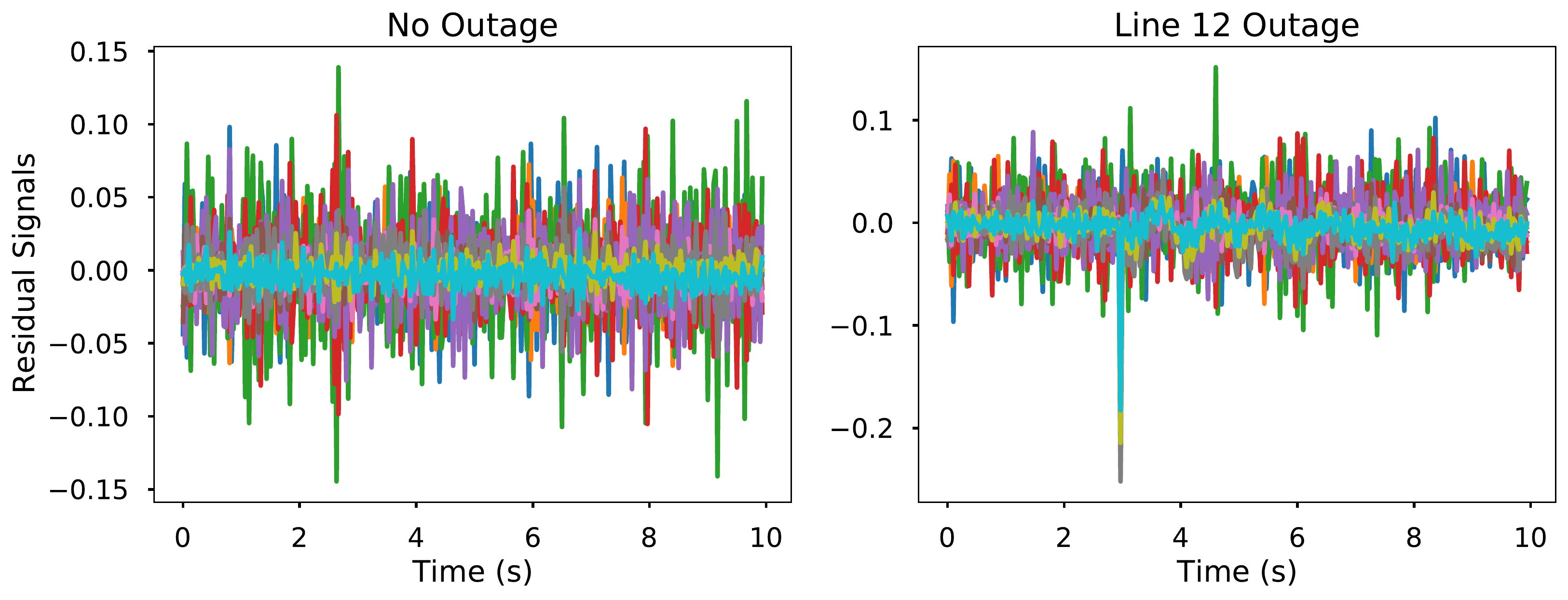}
\caption{\textit{Comparison of the residual signals with no outage and with line 12 outage. A subset of residual signals significantly deviated from the normal mean level and exhibited strong non-Gaussian oscillations after the outage.}}
\label{fig:signals_comparison}
\end{figure}

We propose a real-time detection scheme that utilizes the output function of the SSM detailed in the previous section. In particular, we focus on monitoring the residual between estimated and measured system states. Under an outage-free scenario, the active power generated, transmitted, and consumed in the network are expected to be balanced with only small random load demand fluctuations. Therefore, $\Delta\textbf{P}_{l, k}$, which represents the instantaneous changes in power demand, is assumed to be normally distributed with mean zero under the baseline condition:
\begin{equation}
\label{eqn:normal_distribution}
\Delta\textbf{P}_{l, k} = 
\Delta\textbf{P}_{g, k} - \boldsymbol{Y}_{k} \sim N(\mathbf{0}, \sigma^2\mathbf{I}) \,.
\end{equation}
After a line outage, there are two ways that the above relationship will be violated. First, the outage-induced topology change means that the line admittance of the tripped line becomes effectively zero; the bus admittance matrix thus changes to reflect the post-outage system topology. Therefore, the outage-free AC power flow equation (\ref{eqn:ac_pf}) used to compute the net active power is no longer valid. Thus $\boldsymbol{Y}_{k}$ in (\ref{eqn:normal_distribution}) does not represent the actual net active power changes anymore. Second, the outage triggers a period of transient re-balancing in the system where the generators respond to the sudden power imbalance. The immediately affected buses also experience an abrupt change in the net active power due to the outage. As a combination of these effects, the relationship (\ref{eqn:normal_distribution}) will be violated. For example, using simulation from the IEEE 39-bus system, Fig. \ref{fig:signals_comparison} shows the signals from a normal system and that with an outage at the third second. 

Therefore, the early outage detection problem is formulated as a multivariate process monitoring problem. The multivariate residual signal's significant deviation from the expected distribution indicates an abnormal event, e.g., an outage. For its robustness to non-Gaussian data, superior performance on small to median shifts, and easy of implementation, the multivariate exponentially weighted moving average (MEWMA) control chart, initially developed by \cite{lowry1992multivariate}, is adopted for the detection task. The MEWMA control chart uses an intermediate quantity, $\boldsymbol{Z}_k$, that captures both the current and past signal information from the system. The quantity can be shown to be the weighted average of all past signals with geometrically declining weights. In particular, it can be constructed with the latest SSM output, $\boldsymbol{Y}_k$, as follows:
\begin{equation}
\label{eqn:ewma_z}
\boldsymbol{Z}_k = \lambda (\Delta\textbf{P}_{g, k} - \boldsymbol{Y}_{k}) + (1 - \lambda) \boldsymbol{Z}_{k-1} \,,
\end{equation}
where $\lambda$ is a pre-defined smoothing parameter that controls the extent of the reliance we would like to put on past information. Also, $0 \le \lambda \le 1$ and $\boldsymbol{Z}_0 = \vb{0}$. The statistic under monitoring is then constructed similar to that of a Hotelling's $T^2$ statistic:
\begin{equation}
\label{eqn:ewma_T}
T^2_k = \boldsymbol{Z}_k^T\Sigma_{\boldsymbol{Z}_k}^{-1}\boldsymbol{Z}_k \,,
\end{equation}
where the covariance matrix is 
$$
\Sigma_{\boldsymbol{Z}_k} = \frac{\lambda}{2 - \lambda}\left[1-(1-\lambda)^{2k}\right]\sigma^2\mathbf{I} \,.
$$
An outage alarm is then triggered when the monitoring statistic crosses a pre-determined threshold, $H$, chosen to satisfy a certain false alarm rate requirement:
\begin{equation}
\label{eqn:control_chart}
D = \inf\lbrace k\ge1 : T^2_k \ge H \rbrace \,.
\end{equation}
$D$ is the stopping time of the proposed outage detection scheme. Suppose the onset time of the outage is denoted by $t_o$, then the difference between the stopping time of the control chart and the onset time is the detection delay, i.e., $D - t_o$. One way of judging the effectiveness of different detection schemes is by comparing their detection delays against various line outage events. An ideal detection scheme is therefore able to detect an outage immediately after it happened, i.e., $D - t_o = 0$. 

For MEWMA control charts, it is possible to specify a requirement on the false alarm rate through the selection of $\lambda$ and $H$. 
One way to specify the detection scheme's false alarm rate is through the so-called average run length under zero state ($ARL_0$), i.e., the average number of signals collected before the above detection threshold is reached under an outage-free scenario. Larger $ARL_0$s correspond to more lenient false alarm requirements, but possibly longer detection delays. It is also known that small $\lambda$ values produce control charts more robust against non-Gaussian distributions and have better detection performance for small to medium shifts \cite{montgomery2007introduction}. Given $\lambda$ and a false alarm requirement $ARL_0$, the detection threshold $H$ can be determined by solving an integral equation of Theorem 2 in \cite{rigdon1995integral}\footnote{The equation can be solved using various numerical algorithms or Markov chain approximation, and it be done offline. Interested readers can refer to \cite{knoth2017arl} for a detailed description of the computation procedure required.}. The selection of the parameter values and their impact on the detection scheme will be presented in the simulation study section.

\begin{IEEEproof}[Remark 1 (Limited PMU Deployment)]
The proposed scheme is applicable under a limited PMU deployment since the signal under monitoring can include only buses with PMUs, i.e., $\Delta\textbf{P}_{g}$ includes only generator buses with PMUs. $\Delta\textbf{P}$ can be calculated for load buses with fully observable neighbor buses. The impact of an unobserved neighbor bus on the computation of the bus net active power is limited to an unknown term, $\text{V}_j \text{Y}_{ij} \cos (\theta_i - \theta_j - \alpha_{ij})$, in the AC power flow equation as the neighbor bus' $\theta_j$ and $\text{V}_j$ are not available. This can be mitigated through a careful selection of the PMU locations. Also, unlike \cite{Chen2016} and \cite{yang2020control}, the proposed detection scheme is effective when most generator buses are monitored, a result corroborated by our simulation study, e.g., see Fig. \ref{fig:detection_delay_distribution}. Also, the number of generator buses is typically much smaller than the total number of buses in a system.
\end{IEEEproof}

\section{Generator State Estimation via Particle Filter}
\label{sec:state_estimation}
In the previous section, a unified framework of real-time system monitoring utilizing post-outage transient dynamics computed from dynamic and algebraic state variables, i.e., active power generated and net active power injection, is described. The premise of the unified framework is the availability of accurate state variables data. While algebraic states can be measured by PMUs, generator states are not directly observable. This section shows how the unobservable states could be reliably estimated online using a particle filter.

Online state estimation typically involves the inference of the posterior distribution of the unobservable states $\boldsymbol{X}_k$ given a collection of measurements $\boldsymbol{y}_{0:k}$, denoted by $\pi(\boldsymbol{X}_k | \boldsymbol{y}_{0:k})$. 
For systems with nonlinear dynamics and possibly non-Gaussian noises, e.g., power system, the posterior distribution is intractable and cannot be computed in closed form.
Extended and unscented Kalman filters have been extensively studied to address the above problem, e.g., \cite{Zhao2017,Wang2012}. However, these methods' effectiveness becomes questionable when the underlying nonlinearity is substantial or when the posterior distribution is not well-approximated by Gaussian distribution. Instead, PF is increasingly used for this task, e.g., \cite{Cui2015}, as it handles nonlinearity well and accommodates noise of any distribution with an affordable computational cost \cite{cappe2007overview}. PFs belong to the family of sequential Monte Carlo methods where Monte Carlo samples approximate complex posterior distributions, and the distribution information is preserved beyond mean and covariance. 

In particular, PF approximates $\pi(\boldsymbol{X}_k | \boldsymbol{y}_{0:k})$ by samples, called particles, obtained via an importance sampling procedure. Each particle is assigned an importance weight proportional to its likelihood of being sampled from the posterior distribution\footnote{This type of PF is also known as the bootstrap filter first proposed in \cite{Gordon1993}. The idea is to use the state transition density as the importance distribution in the importance sampling step. More sophisticated algorithms, such as the guided and auxiliary particle filter could be implemented in the same detection framework proposed here. However, these algorithms are, in general, more difficult to use and interpret. For details, readers can refer to \cite{doucet2009tutorial}.}. PF proceeds in a recursive prediction-correction framework. Assuming at time $k$ we have the particles and weights obtained from the previous time step, $\{(\boldsymbol{x}_{k-1}^{i}, w_{k-1}^{i})\}_{1\leq i \leq n}$, where $n$ is the number of particles, the posterior distribution at time $k-1$ is approximated by weighted Dirac delta functions as
\begin{equation}
\pi(\boldsymbol{X}_{k-1} | \boldsymbol{y}_{0:k-1}) \approx \sum_{i=1}^{n} w_{k-1}^{i} \cdot \delta(\boldsymbol{X}_{k-1}-\boldsymbol{x}_{k-1}^{i}) \,,
\end{equation} where $\delta (\cdot)$ is the Dirac delta function, and the weights are normalized such that $\sum_{i=1}^{n} w_{k-1}^{i} = 1$. The algorithm starts by propagating particles from time $k-1$ to time $k$ through the state transition function in (\ref{eqn:de_discrete}), i.e., the prediction step, to obtain the new particles $\{\boldsymbol{x}_k^{i}\}_{1\leq i \leq n}$. The predicted states then have a distribution approximated by 
\begin{equation}
\label{eqn:prediction}
\pi(\boldsymbol{X}_k | \boldsymbol{Y}_{0: k-1}) \approx \sum_{i=1}^{n} w_{k-1}^{i} \cdot \delta(\boldsymbol{X}_k-\boldsymbol{x}_k^{i}) \,.
\end{equation}
When the new measurement $\boldsymbol{y}_k$ arrives, the above approximation is corrected by updating the particles' weights proportional to their conditional output likelihood to obtain the posterior distribution as 
\begin{equation}
\label{eqn:particle_approx}
\pi(\boldsymbol{X}_k | \boldsymbol{Y}_{0: k}) \approx \sum_{i=1}^{n} w_k^{i} \cdot \delta(\boldsymbol{X}_k-\boldsymbol{x}_k^{i}) \,,
\end{equation} where 
$w_k^{i} \propto  w_{k-1}^{i} \cdot  g(\boldsymbol{y}_k | \boldsymbol{x}_k^{i})$.
The intuitive interpretation is that the particles are reweighted based on their compatibility with the actual system measurement. The approximation of the posterior distribution by these particle-weight pairs is consistent as $n \rightarrow +\infty$ at a standard Monte Carlo rate of $\mathcal{O}(n^{-1/2})$ guaranteed by the Central Limit Theorem \cite{doucet2009tutorial}. 

A well-known problem of PF is that the weights will become highly degenerate overtime. In particular, the density approximation will be concentrated on a few particles, and all the other particles carry effectively zero weight. A common way to evaluate the extent of this degeneracy is by using the so-called Effective Sample Size (ESS) criterion \cite{liu2008monte}:
\begin{equation}
\text{ESS}=\left(\sum_{i=1}^{n}\left(w_k^{i}\right)^{2}\right)^{-1} \,.
\end{equation}
In the extreme case where one particle has the weight of 1 and all others of 0, ESS will be 1. On the other hand, ESS is $n$ when every particles has an equal weight of $n^{-1}$.
A resampling move can be used to solve the degeneracy problem where particles with higher weights are duplicated and others removed, thus focusing computational efforts on regions of higher probability. The systematic resampling method is used in our PF as it usually outperforms other resampling algorithms \cite{doucet2009tutorial}. When ESS falls below a threshold, typically $n/2$, we resample $n$ particles from the existing ones. The number of offspring, $n_k^{i}$, is assigned to each particle $\boldsymbol{x}_k^i$ such that $\sum_{i=1}^{n}n_k^{i} = n$. The systematic sampling proceeds as follows to select $n_k^{i}$. A random number $U_{1}$ is drawn from the uniform distribution $\mathcal{U}\left[0, {n}^{-1}\right]$. Then we obtain a series of ordered numbers by $U_{i}=U_{1}+\frac{i-1}{n}$ for $i=2, \ldots, n$. $n_k^{i}$ is the number of $U_{i} \in(\sum_{s=1}^{i-1} w_{s}, \sum_{s=1}^{i} w_{s}]$ where $\sum_{s=1}^{0} w_{s} := 0$ by convention. Finally, resampled particles are each assigned an equal weight $n^{-1}$ before a new round of prediction-correction recursion begins. The detailed PF algorithm with the resampling move is summarized in Algorithm \ref{alg:particle_filter}.
\begin{algorithm}
\caption{Particle Filter for Generator State Estimation}\label{alg:particle_filter}
\begin{algorithmic}[1]
\For{$i = 1, \dots, n$}  \Comment{Initialization}
\State Sample $\tilde{\boldsymbol{x}}_0^{i} \sim \pi_0(\boldsymbol{X})$.
\State Compute initial importance weight
$
\tilde{w}_0^{i} = g(\boldsymbol{y}_0 | \tilde{\boldsymbol{x}}_0^{i})
$
by output function (\ref{eqn:ae_discrete}).
\EndFor
\For{$k \ge 1$}
\If{$\text{ESS} \le n/2$} \Comment{Systematic resampling}
\State Draw $U_{1} \sim \mathcal{U}\left[0, {n}^{-1}\right]$ and obtain $U_{i}=U_{1}+\frac{i-1}{n}$ for $i=2, \ldots, n$.
\For{$i = 1, \dots, n$}
\State Obtain $n_k^{i}$ as the number of $U_i$ such that
$$
U_i \in \left(\sum_{s=1}^{i-1} w_{s}, \sum_{s=1}^{i} w_{s}\right] \,.
$$ 
\State Select $n$ particle indices $j_i \in \{1, \dots, n\}$ according to $n_k^{i}$.
\State Set $\boldsymbol{x}_{k-1}^{i} = \tilde{\boldsymbol{x}}_{k-1}^{j_i}$, and $w_{k-1}^{i} = 1/n$.
\EndFor
\Else 
\State Set $\boldsymbol{x}_{k-1}^{i} = \tilde{\boldsymbol{x}}_{k-1}^{i}$ for $i=1, \ldots, n$.
\EndIf
\For{$i = 1, \dots, n$}
\State Propagate particles \Comment{Prediction}
$$
\tilde{\boldsymbol{x}}_k^{i} \sim f(\boldsymbol{X}_k|\boldsymbol{x}_{k-1}^{i})
$$
via system function (\ref{eqn:de_discrete}).
\State Update weight \Comment{Correction}
$$
\tilde{w}_k^{i} = {w}_k^{i} \cdot g(\boldsymbol{y}_k | \tilde{\boldsymbol{x}}_k^{i}) \,.
$$
\EndFor
\State Normalize weights
$$
w_k^{i} = \frac{\tilde{w}_k^{i}}{\sum_{k=1}^{n}\tilde{w}_k^{k}} \,, \text{for } i = 1, \dots, n \,.
$$
\EndFor
\end{algorithmic}
\end{algorithm}

\begin{IEEEproof}[Remark 2 (Unknown System Parameter Estimation)]
We have assumed that system parameters in the power system SSM are known and static; therefore, the PF's state estimation is reliable. In real-world applications, these parameters may be known but slow-varying due to factors like system degradation. While parameter estimation in a nonlinear system is generally difficult and outside the scope of this work, there is a natural extension from the particle filtering framework to tackle the problem. An online expectation maximization (EM) algorithm based on the particles can learn the parameters as data arrives sequentially. The EM algorithm is an iterative optimization method that finds the maximum likelihood estimates of the parameters in problems where unobservable variables are present. It can be reformulated to perform the estimation online using the so-called sequential Monte Carlo forward smoothing framework when the complete-data density, i.e., $p_\vartheta(\boldsymbol{X}_{0:k}, {\boldsymbol{Y}}_{0:k})$ where $\vartheta$ denote the set of unknown parameters, is from the exponential family \cite{yildirim2013online}.
\end{IEEEproof}

\section{Simulation Study}
\label{sec:results}
\subsection{Simulation Setting}
The proposed PF-based outage detection scheme is tested on the IEEE 39-bus 10-machine New England system~\cite{athay1979practical}. System transient responses after an outage are simulated using the open-source dynamic simulation platform COSMIC \cite{Song2016}. The simulation results are assumed to be the true generator states, and corrupted measurements are synthesized from the noise-free simulation data. Ten PMUs are assumed to be installed at bus 19, 20, 22, 23, 25, 33, 34, 35, 36, and 37, covering five generator buses and their connected load buses. Their sampling frequency is assumed to be 30 samples per second. Each simulation runs for 10 seconds, and the outage happens at the third second. A line outage is detected if the monitoring statistic crosses the detection threshold by the end of the simulation. The detection thresholds of all schemes presented are selected by satisfying a false alarm constraint of 1 in 30 days. The global constants are $f_0 = 60 $ Hz and $\omega_s = 1.0 $ p.u.. For the SSM, state function noise $\epsilon_k$ are assumed to be uncorrelated and homogeneous with a standard deviation of $0.01\% \cdot \text{P}_{g,k}$ in (\ref{eqn:de_discrete}). Output function error $\boldsymbol{\eta}_k$ are assumed to follow a zero-mean Gaussian distribution with a standard deviation of 
$1\% \cdot (\text{P}_{g, k}- \text{P}_{k})$ in (\ref{eqn:ae_discrete}). 

\subsection{Illustrative Outage Detection Example}  
\label{sec:results:example}
\begin{figure}[!t]
\centering
\includegraphics[width=1\linewidth]{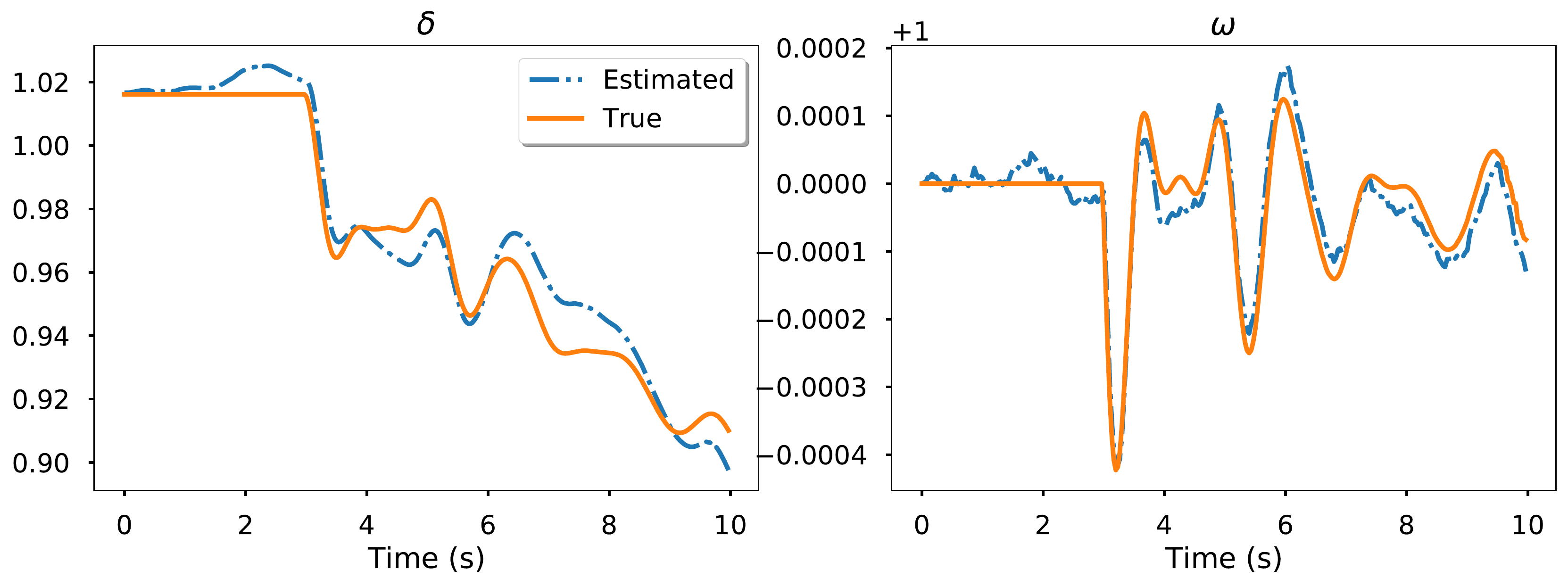}
\caption{\textit{State estimation result of the particle filter on $\delta$ and $\omega$ of Bus 33. The algorithm can estimate $\omega$ accurately, while the estimation of $\delta$ has biases after the outage. The changes in $\delta$ are sufficiently captured, which are more critical for the detection scheme.}}
\label{fig:line_18_state_estimation_eg}
\end{figure}

We present line 11 outage to illustrate the working of the detection scheme. Fig. \ref{fig:line_18_state_estimation_eg} shows a typical performance of the particle filter used to estimate generator bus states. The rotor angular speed, $\omega$, can be accurately tracked while the rotor angular position, $\delta$, has some biases after the outage. This is acceptable since the focus is on capturing the abnormal changes, i.e., $\Delta\delta$ and in turn $\Delta P_g$, in response to the outage rather than accurate state estimations.

\begin{figure}[!t]
\centering
\includegraphics[width=1\linewidth]{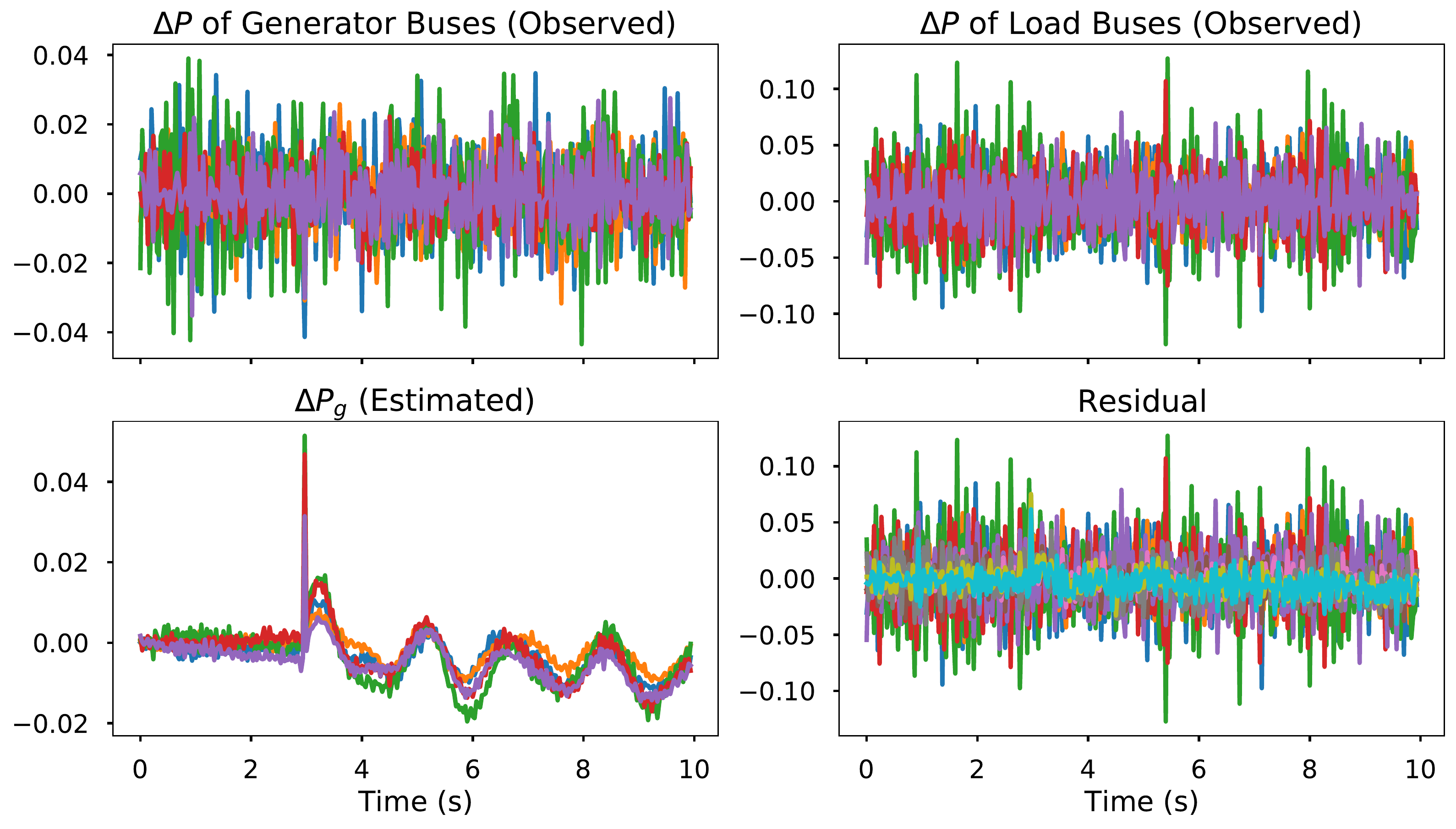}
\caption{\textit{Output signals of the detection scheme for line 11 outage. Each line represents data from a bus equipped with a PMU. Abnormal disturbances in generator rather than load buses contributed to early detection in this case.}}
\label{fig:detection_signal_breakdown}
\end{figure}

\begin{figure}[!t]
\centering
\includegraphics[width=1\linewidth]{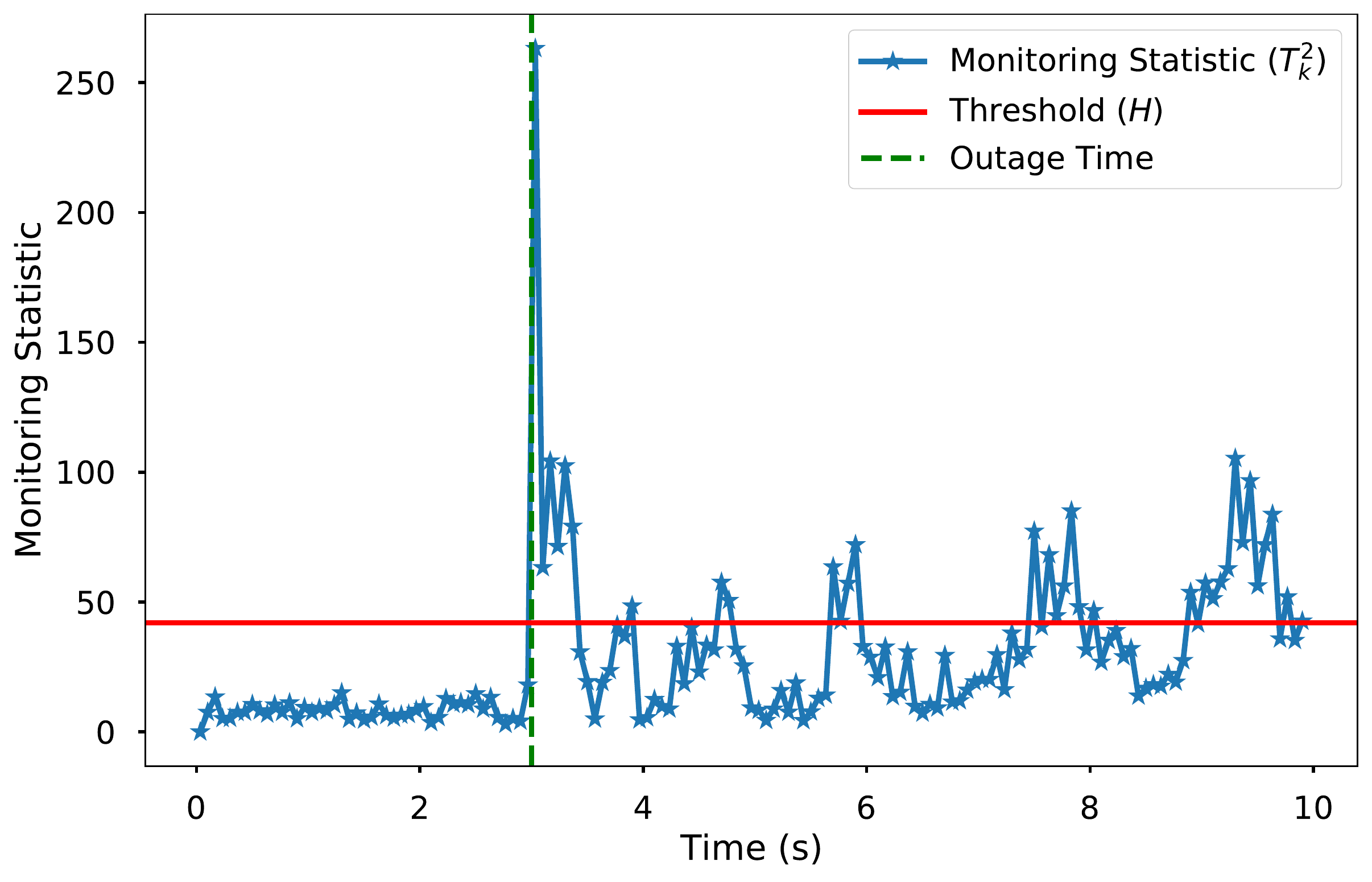}
\caption{\textit{Progression of MEWMA monitoring statistic for detecting line 11 outage. After the outage, the monitoring statistic crosses the detection threshold immediately and remains high afterward. The outage is successfully detected with no detection delay. Points are downsampled to half for clarity}}
\label{fig:line_11_monitor_statistics}
\end{figure}

One advantage of the proposed scheme is the ability to break down monitored signals and pinpoint the channels leading to a detection. Fig. \ref{fig:detection_signal_breakdown} shows such a breakdown. The upper two channels are the observable net active power information, and the lower-left one is the estimated generator information. They register different signal strength levels depending on the outage location, e.g., the magnitude of initial shock and the transient oscillation duration. The proposed scheme can detect outages as long as one of them picks up significant changes. It is clear in this case that the signals from PMU measurements do not contribute meaningfully to the detection. Instead, the changes in generated active power  on generator buses display significant abnormal fluctuations, leading to the outage detection. 
The typical progression of the monitoring statistic, $T_{k}^2$, computed via MEWMA from the output signals is shown in Fig. \ref{fig:line_11_monitor_statistics}. Before the outage, the statistic remains close to zero. After the outage at the third second, it increases rapidly and crosses the threshold. Thus, the scheme raises an outage alarm, and no detection delay is incurred. 

\subsection{Results and Discussion} 
This section presents the result computed from 1000 random simulations of each line outage. It includes the comparison with other state-of-the-art methods, the detection delay over all outage scenarios, the effect of the outage location as well as the impact of the smoothing parameter $\lambda$.

\subsubsection{Comparison with Other Methods}
The proposed method's performance is compared with three other methods in Table \ref{tab:delay_comparison_39}. The presented line outages are more difficult to detect. The first method for comparison is based on the DC power flow model from \cite{Chen2016} and the second based on subspace identification from \cite{Hosur2019}. Both of them are tested using a full PMU deployment. The third is based on the AC power flow model where 10 PMUs are assumed to be available \cite{yang2020control}. The proposed method is under the column of ``Unified". The average detection delays and their standard deviations are presented; A dash means a missed detection. While the DC method detected immediately outages at line 11 and 15, even under a full PMU deployment, it missed half of the selected outages. The  difference seems to suggest that the DC method depends on a strong initial signal. The subspace and the AC method can detect all outages. However, they incurred longer delays compared to the unified scheme. This might be because only algebraic states are used in their monitoring. 

\begin{table}
\begin{threeparttable}
\caption{Detection Delay (s) by Different Detection Schemes}
\label{tab:delay_comparison_39}
\centering
\begin{tabular}{rllll} 
\hline
\hline
\multicolumn{1}{l}{}     & \multicolumn{4}{c}{Average Detection Delay\tnote{1}}  \\ \hline
\multicolumn{1}{c}{Line} & DC - full & Subspace - full  & AC &  Unified \\ \hline
2 & 1.165 (0.006)   & 2.822 (1.924) & 0.283 (0.263) & \textbf{0.012} (0.183) \\
6 & \textendash     & 3.060 (2.011) & 0.246 (0.129) & \textbf{0.052} (0.463) \\
11& \textbf{0} (0)  & 3.048 (1.969) & 0.602 (0.205) & 0.058 (0.077) \\
15& \textbf{0} (0)  & 2.634 (1.850) & 0.005 (0.034) & \textbf{0} (0) \\
19& \textendash     & 2.836 (2.018) & 0.335 (0.378) & \textbf{0.160} (0.315) \\
26& \textendash     & 2.850 (1.958)  & 0.385 (0.228) & \textbf{0} (0)      \\
\hline          
\end{tabular}
\begin{tablenotes}
\item [1] Standard deviation appears in parentheses.
\end{tablenotes}
\end{threeparttable}
\end{table}

\begin{figure}[!t]
\centering
\includegraphics[width=1\linewidth]{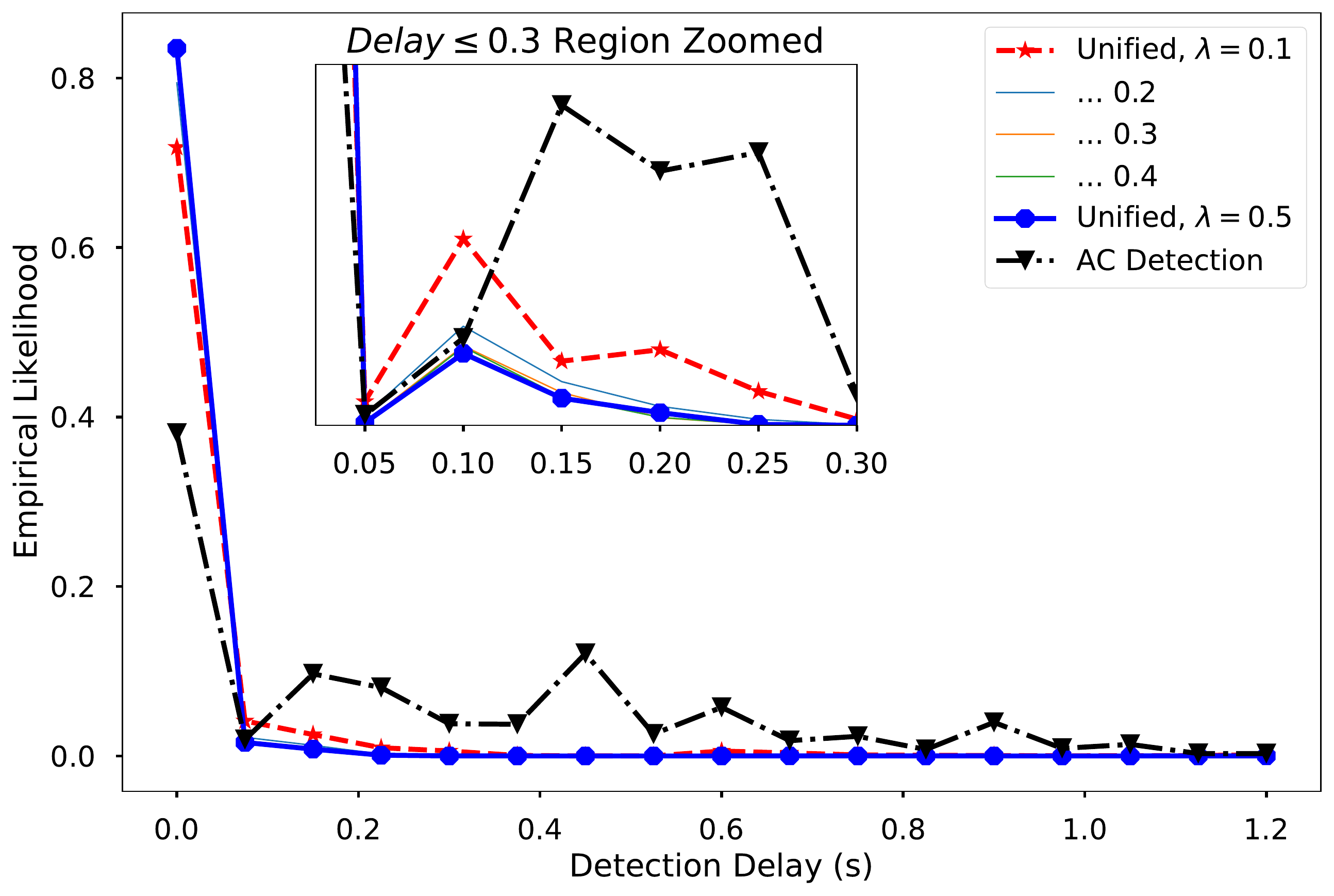}
\caption{\textit{Comparison of the empirical likelihood of detection delays (s) for the proposed and the AC power flow equations-based scheme. The proposed scheme has a higher percentage of zero detection delays. It can detect most outages within 0.2 seconds and the AC detection scheme in one second.}}
\label{fig:detection_delay_distribution}
\end{figure}

The AC scheme comes close to the proposed scheme in terms of coverage and delay. However, the proposed scheme is still faster and, as discussed in the next section, improves substantially on the robustness to relative locations to available PMUs. The empirical likelihood of detection delays over all simulated line outages is presented in Fig. \ref{fig:detection_delay_distribution}. Intuitively, the scheme is faster at detecting outages when the area under the curve towards the left of the figure is larger. In this case, the proposed scheme has a much higher chance of detecting outages with zero detection delay than the AC scheme. For example, the best-performing scheme ($\lambda = 0.5$) detects most outages within 0.2 seconds, whereas the AC scheme detects most outages within 1 second.

\subsubsection{Effect of Outage Location Relative to the PMUs}
\begin{figure}[!t]
\centering
\includegraphics[width=1\linewidth]{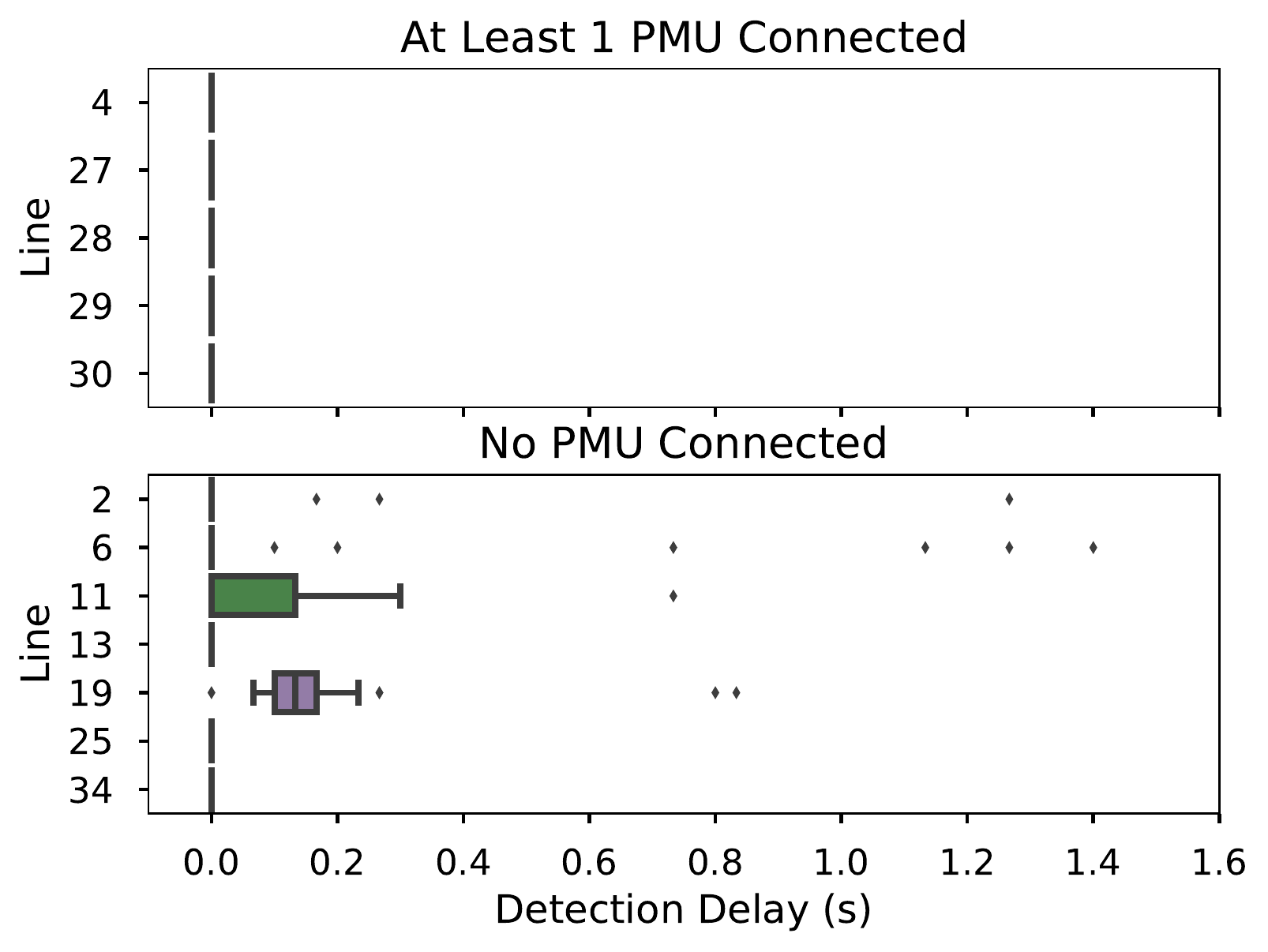}
\caption{\textit{Box plot of the empirical distributions of detection delays in seconds for lines with at least 1 PMU nearby and those without a PMU.}}
\label{fig:boxplot_delay_pmu}
\end{figure}

Significant variations of detection delays was previously reported for outages at different lines relative to the PMU locations \cite{yang2020control}. Fig. \ref{fig:boxplot_delay_pmu} shows a comparison of detection delays for outage lines with at least one PMU connected to it versus those with no PMU nearby. Since only ten buses are equipped with PMUs, most lines belong to the second group\footnote{All line outages not displayed can be detected with zero mean detection delay, except for line 35 and 36, which are often undetected.}. While outages at line 11 and 19 are often detected with 0.1-second delay, most outages are detected immediately regardless of the relative position to the PMUs. Line 11 connects to the slack bus, and its outage creates a minimal disturbance in all three output channels. This result demonstrates the spatial advantage of the proposed method and its robustness to the outage locations.

\subsubsection{Effect of $\lambda$ on Detection Delay and Rate}
\begin{figure}[!t]
\centering
\includegraphics[width=1\linewidth]{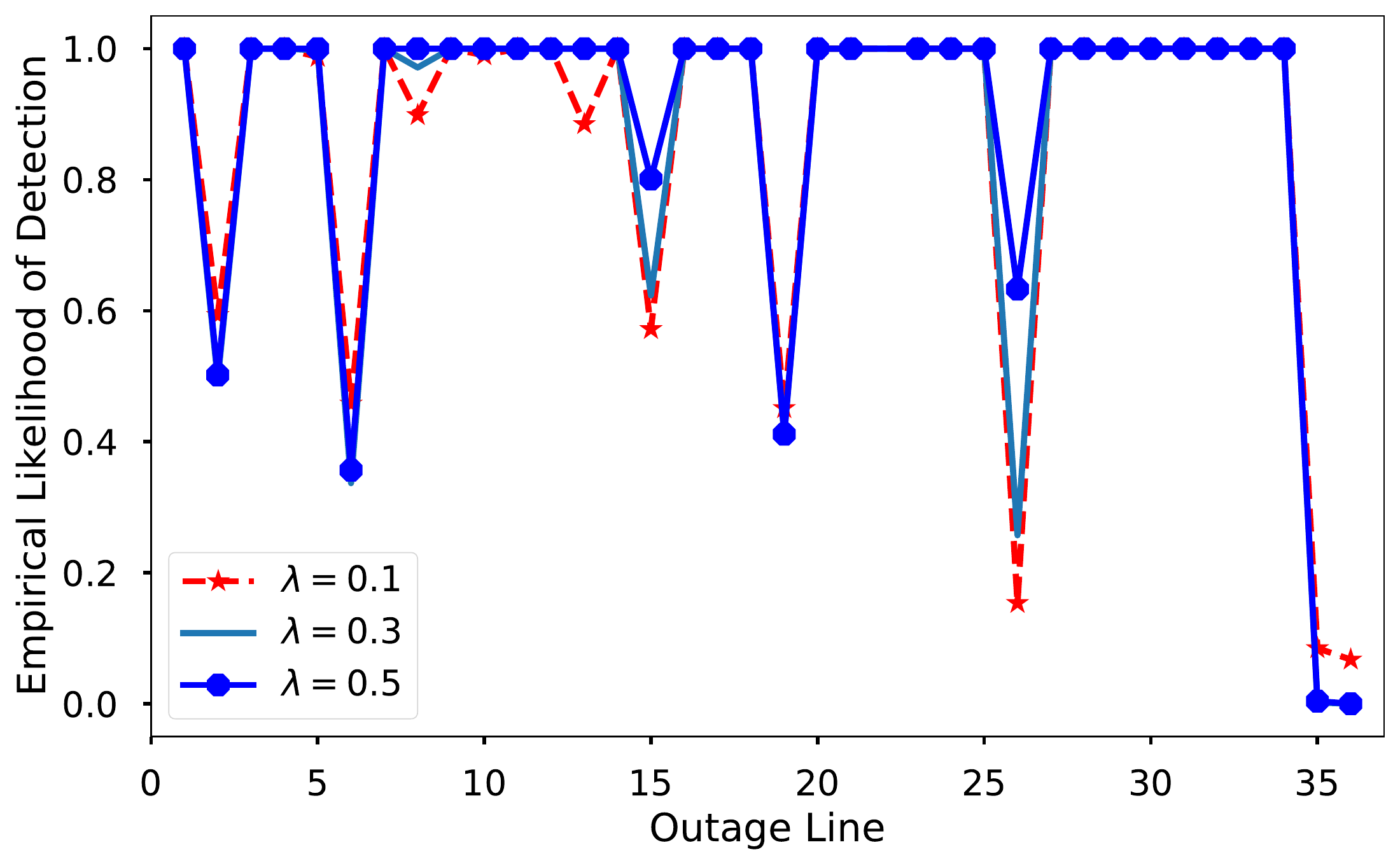}
\caption{\textit{Empirical detection likelihood for simulated outages under different $\lambda$s. While 28 out of the 35 line outages are detected with over 90\% likelihood, a small group of outages is difficult to detect regardless of the $\lambda$ value.}}
\label{fig:detection_rate}
\end{figure}

The smoothing parameter $\lambda$'s impact on the detection performance can be seen from two aspects: detection delay and detection rate. Fig. \ref{fig:detection_delay_distribution} shows the delay likelihood with different $\lambda$ values. The scheme with a larger $\lambda$ value tends to record higher proportion of zero-delay detection. This is likely due to the higher weight given to the current measurement. 
On the other hand, Fig. \ref{fig:detection_rate} shows the empirical likelihood of successful detection for all 35 line outages over 1000 simulations. For all values of $\lambda$, the scheme can detect 28 out of 35 outages over 90\% of the time. In some cases, larger values of $\lambda$ tend to have a better detection rate, e.g., line 8, 13, 15, and 26. The reason might be that these outages produce more severe initial shock relative to their after-outage oscillation. Hence, larger values of $\lambda$ help to capture the immediate shock. Also, a small group of outages is challenging to detect regardless of the $\lambda$ value, e.g., line 2, 6, 19, 35, and 36. Diagnostic inspection of these cases' signals reveals that they produced weak system disturbances, especially from the generators, hence not triggering an outage alarm. The weak disturbance might be because these lines are connected to buses that serve near zero loads. 

\section{Conclusion}
\label{sec:conclusion}
We have proposed a unified framework of online transmission line outage detection. The framework utilizes information from both generator dynamic state and algebraic state variables. The signals are obtained through nonlinear state estimation of particle filters and direct measurements of PMUs. They are used for monitoring and outage detection by MEWMA control charts while meeting a pre-specified false alarm constraint. The approach is shown to be quicker at detecting outages and more robust to unknown outage locations under a limited PMU deployment. Further research can be done to improve the detection scheme’s effectiveness by investigating the optimal installation location of limited PMUs. Also, it is observed that a group of lines is consistently challenging to detect regardless of the detection schemes or parameter designs used. More work can to be done in this area so that these detection blind spots could be reduced.



\section*{Acknowledgment}
This work was partially supported by the Singapore AcRF Tier 1 Fund under Grant R-266-000-123-114 and the Future Resilient Systems program at the Singapore-ETH Centre, which was established collaboratively between ETH Zurich and in part by the National Research Foundation (FI 370074011), Prime Minister’s Office, Singapore under its Campus for Research Excellence and Technological Enterprise (CREATE) program.

\ifCLASSOPTIONcaptionsoff
  \newpage
\fi



%

\begin{IEEEbiography}{Xiaozhou Yang}
received the B.S. degree in industrial and systems engineering from the National University of Singapore, Singapore, in 2017. 
He is currently a Ph.D. student in the Department of Industrial Systems Engineering and Management of the same university. He is also a researcher at the Future Resilient Systems program of Singapore-ETH Centre. 
His research interests include advanced data analytics in power system condition monitoring, real-time outage detection, and identification.
\end{IEEEbiography}

\begin{IEEEbiography}{Nan Chen}
received the B.S. degree in automation from Tsinghua University, Beijing, China, in 2006, the M.S. degree in computer science in 2009, and the M.S. degree in statistics and the Ph.D. degree in industrial engineering from the University of Wisconsin-Madison, Madison, WI, USA, both in 2010.
He is currently an Associate Professor with the Department of Industrial Systems Engineering and Management, National University of Singapore, Singapore. His research interests include statistical modeling and surveillance of engineering systems, simulation modeling design, condition monitoring, and degradation modeling.
\end{IEEEbiography}

\begin{IEEEbiography}{Chao Zhai (S'12-M'14)} received the Bachelor's degree in automation engineering from Henan University in 2007 and earned the Master's degree in control theory and control engineering from Huazhong University of Science and Technology in 2009. He received the PhD degree in complex system and control from the Institute of Systems Science, Chinese Academy of Sciences, Beijing, China, in June 2013. From July 2013 to August 2015, he was a post-doctoral fellow with the University of Bristol, Bristol, UK. From October 2015 to January 2016, he was a research associate with the University of Hong Kong.
From February 2016 to October 2019, he was a research fellow with Nanyang Technological University and Singapore-ETH Centre, Singapore. 
Since November 2019, He has been a Professor at the School of Automation, China University of Geosciences, Wuhan, China. His current research interests include cooperative control of multi-agent systems, power system stability and control, social motor coordination, and evolutionary game theory.
\end{IEEEbiography}




\end{document}